# INTEGRATED DETECTOR CONTROL AND CALIBRATION PROCESSING AT THE EUROPEAN XFEL

A. Münnich, S. Hauf, B.C. Heisen, F. Januschek, M. Kuster, P.M. Lang, N. Raab,
T. Rüter, J. Sztuk-Dambietz, M. Turcato
European XFEL GmbH, Albert-Einstein-Ring 19, 22761 Hamburg, Germany

*Abstract*

The European X-ray Free Electron Laser is a high-intensity X-ray light source currently being constructed in the area of Hamburg, that will provide spatially coherent X-rays in the energy range between 0.25 keV and 25 keV. The machine will deliver 10 trains/s, consisting of up to 2700 pulses, with a 4.5 MHz repetition rate. The LPD, DSSC and AGIPD detectors are being developed to provide high dynamic-range Mpixel imaging capabilities at the mentioned repetition rates. A consequence of these detector characteristics is that they generate raw data volumes of up to 15 Gbyte/s. In addition the detector's on-sensor memory-cell and multi-/non-linear gain architectures pose unique challenges in data correction and calibration, requiring online access to operating conditions and control settings. We present how these challenges are addressed within XFEL's control and analysis framework Karabo, which integrates access to hardware conditions, acquisition settings (also using macros) and distributed computing. Implementation of control and calibration software is mainly in Python, using self-optimizing (py) CUDA code, numpy and iPython parallels to achieve near-real time performance for calibration application.

## INTRODUCTION

The European X-ray Free Electron Laser (XFEL.EU) is an international research facility currently under construction in the area of Hamburg, Germany, which will start operation at the end of 2016 [1]. The superconducting linear accelerator of the facility will deliver electron bunches with an energy of up to 17.5 GeV, arranged in trains of typically 2700 bunches at a repetition rate of 4.5 MHz. Each train will be followed by a gap of 99.4 ms, during which data is read out from the detector front-end. The X-ray pulses will be particularly intense and ultra-short, down to a pulse length of less than 100 fs with a peak brilliance of $10^{33}$ photons/s/mm$^2$/mrad$^2$ per 0.1 % bandwidth [2].

The high repetition rate and the high peak brilliance pose unique challenges in terms of performance and radiation tolerance on the X-ray detector systems employed at the facility. As a result, each scientific instrument uses a detector optimized for its specific needs in terms of energy range, resolution, operation conditions, etc. Three large 2D area imaging detectors are being developed: the Adaptive Gain Integrating Pixel Detector AGIPD [3], DePFET Sensor with Signal Compression DSSC [4, 5] and a Large Pixel Detector LPD [6], which can run at the high repetition rate of 4.5 MHz. Smaller detectors like pnCCDs [7, 8] and Fast CCDs [9, 10] can be operated at 10 Hz. In addition, detector technologies such as Silicon Drift Detectors (SDDs), Micro-Channel Plates (MCPs) or silicon strip detectors [11] are being evaluated for specialized applications.

The large data volumes produced by the MHz-rate 2D detectors (10 to 15 GB per second per detector) and the huge number of calibration parameters (of the order of $10^9$) require new concepts in data handling, calibration and processing in an integrated manner, aspects of which we will present.

## THE KARABO DETECTOR CONTROL SOFTWARE FRAMEWORK

The software framework Karabo [12] will be used to control and manage the photon beamlines at the European XFEL as well as components within the experimental hutches including the detectors. Karabo's generic functionality covers all aspects from data acquisition (DAQ) over equipment control to data analysis. Figure 1 illustrates the different component categories available in the framework and their interaction via a JMS message broker or direct peer-to-peer (p2p) interfaces. In the following we present examples of the usage in the aforementioned categories: the control of a pnCCD-detector setup including the associated vacuum systems employing device composition, the control, DAQ and data-processing of an SDD-detector as well as pipeline-data analysis in the context of detector data correction and calibration.

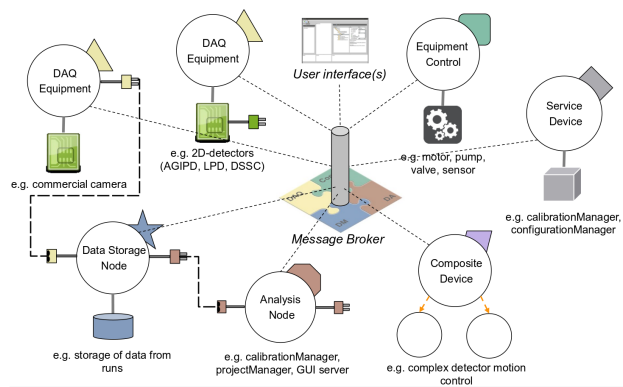

Figure 1: Schematic overview of Karabo's design concept. Functionality is implemented into the framework as devices (either `C++` or `Python3`), which can communicate either via a JMS message broker (shown here as monolithic, but in fact supporting dynamic fail-over and clustering) or p2p interfaces.

## Control and Composition – the pnCCD Setup

The pnCCD detector operates in vacuum and at low temperatures, in order to be able to image X-rays at energies down to 50 eV. For system safety venting has to occur in a controlled fashion and it has to be insured that the device is warmed prior to doing so. Figure 2 shows the control GUI for the vacuum and cooling system of the pnCCD setup which consists of several valves, pumps and a chiller. The status of each component is indicated by its color. User interaction is realized using buttons and relevant information about temperature and pressure is shown. The buttons shown next to components in the GUI are meant for manual interaction, i.e. when they are controlled through the respective Karabo device individually. More complex procedures like pumping to nominal operating conditions, venting or cooling of the system can be implemented either as macros or composite devices on top of this manual control. Figure 3 shows the UML activity diagram, which describes the program flow the "pump" button starts. Several devices are involved relying on conditions set from values from Karabo-controlled pressure and temperature sensors. Such diagrams are created for each procedure and document program states, error conditions and flow.

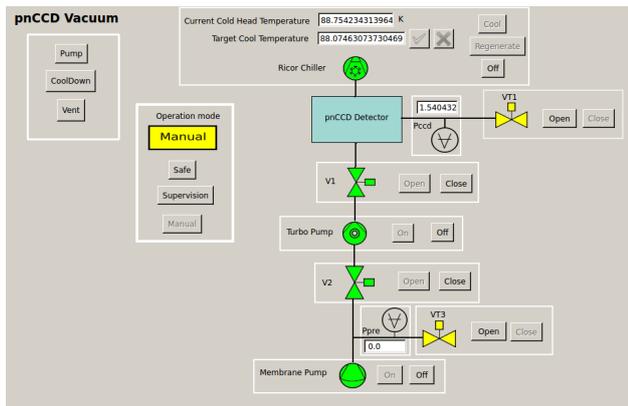

Figure 2: The control GUI for the vacuum and cooling system of the pnCCD setup, laid out as a functional diagram.

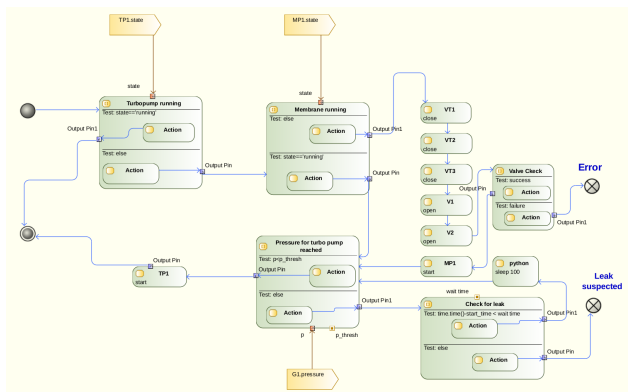

Figure 3: UML activity diagram of the pumping procedure for the pnCCD setup.

## Detector Control, DAQ and Data Processing of an SDD Detector

Silicon drift detectors (SDDs) are 1D detectors used at XFEL.EU as reference detectors for characterization of X-ray sources and for cross-calibration with 2D detectors. The control and readout of an Amptek Fast-SDD detector [13], via Ethernet, has been integrated into Karabo as a device. Through Karabo macros this allows the SDD to be used in automated procedures: e.g taking data at intervals determined by a positional scan using linear stages. Figure 4 shows the corresponding Karabo GUI; in addition to monitoring and control the acquired spectra can be visualized, written into an HDF5 file or streamed via a p2p connection to be processed by another device. As currently implemented, such a "compute" device monitors this output channel, calibrates the incoming data and finds peaks in the spectra in order to identify chemical elements by their fluorescence emissions.

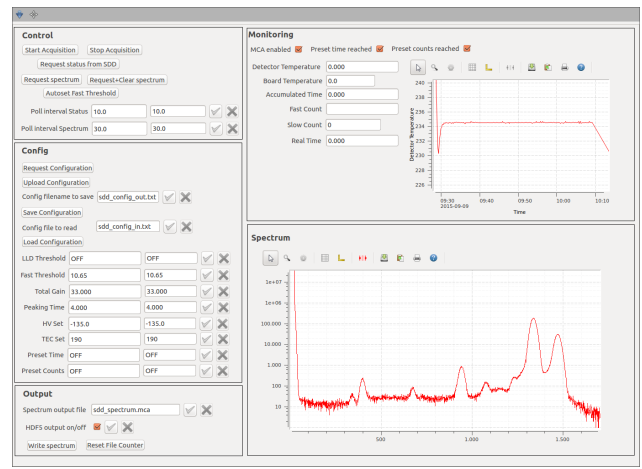

Figure 4: The Karabo GUI for controlling and monitoring an AmpTek Fast SDD detector.

## Python-based Calibration and Data Analysis Suite

European XFEL facility users will not interact with raw detector data, which is the facilities main archival product. Instead, calibrated data sets will be the main user-accessible data product, and calibration and correction will be facility-provided [14]. The data-flow from detector acquisition to end-user availability is illustrated in Fig. 5.

Calibration of the MHz-rate 2D detectors poses a challenge due to the large number of calibration constants. For the non-linear gain calibration of the 1 Mpixel DSSC detector [15] these for example amount to about $10^9$, with similar numbers for LPD and AGIPD where per-pixel, per-gain and per-memory cell calibration is required. The diverse range of experiments, where some have single photon events with low event rates and others densely filled images mandate different detector operation modes, which may need (partially) different calibration data.

In order to manage calibration-associated information, which may be produced at intervals of per-run to a few months, a MySQL database has been developed, which interacts with dedicated python devices through RESTful interfaces. It is able to resolve calibration constants by detector operating condition, maintain a set of current calibration data and associated validity periods thereof, as well as keep track of detector alignment information.

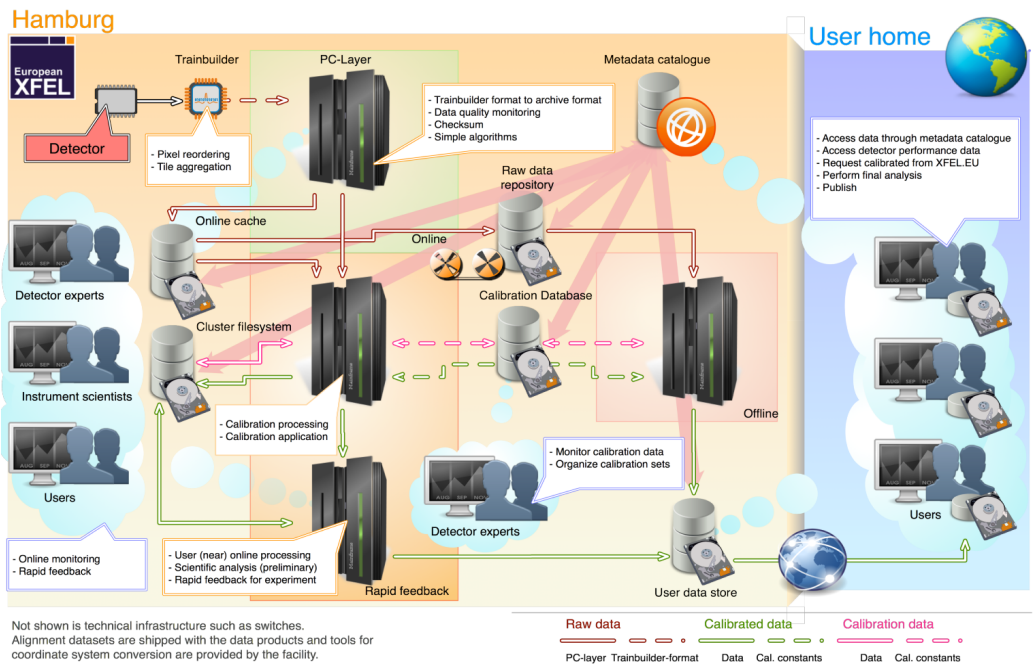

Figure 5: The data processing chain at the European XFEL. Raw data from the detectors (red arrows) is persisted and (selectively) forwarded via Karabo p2p interface to online calibration. Calibrated data (green arrows) is produced either from the online stream or by processing offline data. Calibration data (pink arrows) is processed from the online cache and inserted into the calibration database. In all cases facility users will only access calibrated datasets.

The production and application of the calibration and correction constants managed by this database may occur as part of two general scenarios: interactive data analysis and automated, pipeline processing, the latter ideally with near-real-time performance. A python library which provides a common codebase for both applications is being developed. It can be used as part of an iPython (notebook) based interactive analysis [16] or automated in a distributed Karabo system communicating via p2p links. Underlying, numpy [17] and scipy [18] are used for calculations on the CPU, ipcluster for concurrent processing and pyCuda [19] for GPU-accelerated computing. Further optimization using Numba JIT-technology [20] are being investigated and near-real time performance for calibration application has been demonstrated [21].

Figure 6 shows an example of pipelined processing within Karabo. Data acquired by an LPD prototype has been offset- and common mode-corrected on the GPU. In the figure the raw and corrected data are contrasted, making apparent that proper calibration is required to reveal the full diffraction patterns produced by the ferrous solution sampled with monochromatic 18 keV X-rays at the ESRF facility. The required calibration and correction constants were retrieved from the calibration database.

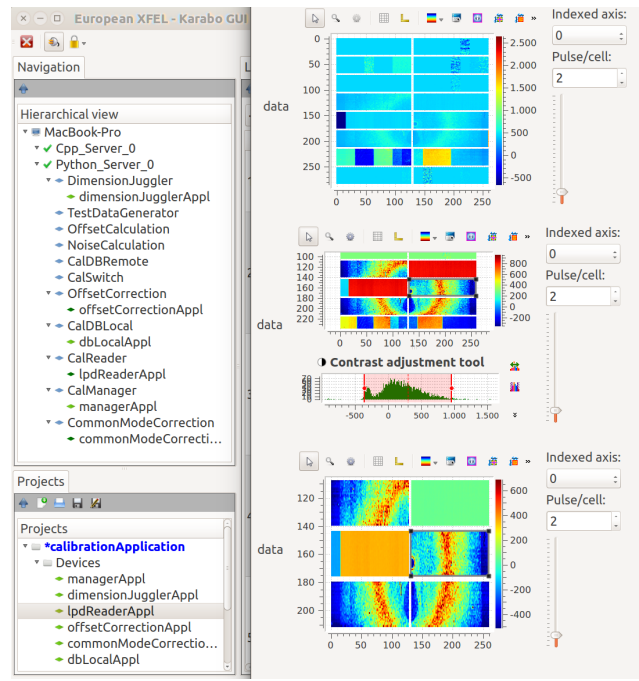

Figure 6: Example of calibration processing within Karabo. On the left the device instances interacting via p2p interfaces are visible. The superimposed panel shows preview images of uncalibrated (top), offset-corrected (center) and offset- and common mode-corrected data showing diffraction by a ferrous solution images by an LPD prototype.

## CONCLUSION

We have presented examples of the integrated control and processing within the Karabo software framework, needed to make optimal use of the capabilities of European XFEL detector systems. The examples range from control and monitoring tasks, an integrated DAQ and control system for an SDD detector, to high-performance calibration processing and associated interaction with a calibration database.

## ACKNOWLEDGMENT

The authors acknowledge the support of the Control and Scientific computing (CAS), Advanced Electronics (AE) and IT and Datamanagement (ITDM) groups at XFEL.EU.